# Machines that test Software like Humans

Anurag Dwarakanath
Accenture Technology Labs
Bangalore, India
anurag.dwarakanath@accenture.com

Neville Dubash
Accenture Technology Labs
Bangalore, India
neville.dubash@accenture.com

Sanjay Podder
Accenture Technology Labs
Bangalore, India
sanjay.podder@accenture.com

*Abstract*—Automated software testing involves the execution of test scripts by a machine instead of being manually run. This significantly reduces the amount of manual time & effort needed and thus is of great interest to the software testing industry. There have been various tools developed to automate the testing of web applications (e.g. Selenium WebDriver); however, the practical adoption of test automation is still miniscule. This is due to the complexity of creating and maintaining automation scripts. The key problem with the existing methods is that the automation test scripts require certain implementation specifics of the Application Under Test (AUT) (e.g. the html code of a web element, or an image of a web element). On the other hand, if we look at the way manual testing is done, the tester interprets the textual test scripts and interacts with the AUT purely based on what he perceives visually through the GUI. In this paper, we present an approach to build a machine that can mimic human behavior for software testing using recent advances in Computer Vision. We also present four use-cases of how this approach can significantly advance the test automation space making test automation simple enough to be adopted practically.

## I. INTRODUCTION

Software testing is often said to take a significant portion of the overall effort in software development with estimates ranging from 30% to 90% [3]. Thus, automation in software testing is often attempted to reduce the amount of manual effort needed. A popular option is the automatic execution of test scripts. A test script is a detailed set of 'instructions' to exercise a particular path in the Application Under Test (AUT). An example of a test script is shown in Figure 1. The test script generates flight options between two cities on a travel booking portal.

| Test Script – 1 | Step 1 | Open browser and go to www.expedia.com |
|---|---|---|
| | Step 2 | Enter 'New York' into textbox 'Flying From' |
| | Step 3 | Enter 'Madrid' into the textbox 'Flying to' |
| | Step 4 | Click on the 'Search' button. Verify the page title is `Search Results' |

**Figure 1. An example of a test script.**

A manual tester would read these test steps & perform the appropriate action on the AUT. If we look at step 2, the tester would first locate a textbox called 'Flying From' among the various web elements displayed on the web page. This is done based on what the tester perceives visually. The tester then enters the text 'New York' into this textbox.

Test automation aims to execute the test scripts automatically without manual intervention. Existing automation tools include Selenium WebDriver [4] and Sikuli [5]. To automate test scripts through these tools, specific code needs to be written. For example, step 2 of Figure 1 can be automated in Selenium WebDriver as shown in Figure 2.

```
WebElement FlyingFrom =
driver.findElement(By.id("src-txt"));

FlyingFrom.sendKeys("New York");
```

**Figure 2. The automation code uses implementation specifics of the AUT (value of the html attribute of 'id').**

There are two problems with the current techniques of test automation – a) the creation of test scripts requires information of the underlying implementation specifics of the AUT. In the example of Figure 2, the textbox is identified by the value of the html attribute 'id' (which the developer of the AUT has given a value of "src-txt" in the example). For a tester to author this automation step, he needs to not only know the programming language (Java in the above example), but also needs to look through the html code of the AUT and identify the code fragment that implements the textbox. b) Once the automation script has been created, maintenance of the script takes significant effort as well, since any change in the implementation of the AUT would require corresponding changes to the automation script. For example, if the value of the 'id' attribute changes in a new version of the AUT or the textbox is rendered by a new framework, the automation script needs to be updated. There are other cases leading to creation and maintenance issues as well, which we will review in Section III.

The problems in creation and maintenance are so severe in practice that projects often find low or no benefits from the adoption of test automation [6] [7].

The fundamental reason for the problems in creation and maintenance is the dependence of the automation script on the implementation specifics of the AUT. On the other hand, manual testing is not prone to these problems even though it goes through the phases of creation and maintenance. This is because, manual test scripts are written in a natural language, such as English, without referring to any implementation specifics (see Figure 1). It is therefore easy to create and maintain. To execute the manual test script, a tester interprets the instructions (such as 'click'), locates the web element (such as 'Flying from textbox') in the GUI of the AUT and performs the action. Importantly, the identification of the web elements is done based on the interpretation of visual content by the tester.

Our goal is to build a machine that can mimic human behavior in the execution of test scripts. The method:
a) Should allow a tester to create an automation script which is as close as possible to English statements. The script should not require any specific implementation details of the AUT
b) Should interpret the statements into actions & web elements. The web elements should be identified from the GUI and the appropriate action should be performed.

To achieve this goal, we have built a Domain Specific Language (DSL) that allows a tester to create test scripts in a simple English-like language. These steps are parsed into actions and web elements. Note that, at this juncture, the web elements are described in textual form. We then leverage a deep neural network to locate the (x, y) co-ordinates of these web elements in the GUI of the AUT (e.g. locate the co-ordinates based on the text of 'Flying from textbox'). Our method then performs the action on the (x, y) co-ordinates.

There have been recent advances in Deep Learning and Computer Vision that give credence to our approach. We have witnessed methods that automatically identify natural objects (such as cats) in images. This has been driven through novel learning methods leveraging deep neural networks in association with large datasets [1]. Results have shown a near human level performance. We look to adapt such techniques for the identification of web elements from the screenshots of the GUI. Further, the variations possible in the way a web element is visually rendered is much smaller than that of natural objects. Thus, we believe that our deep learning based approach can show good accuracy in practice.

To the best of our knowledge, this is the first application of deep neural networks for object recognition in the domain of software testing..

## II. RELATED WORK

The techniques for web automation can be broadly classified into: a) Programming based automation; b) Record & Replay; and c) Image recognition based automation. We will briefly review the first two techniques and review the image recognition method in depth, as cursorily it might appear similar to our proposed approach.

Programming based automation requires a tester to author automation code in a programming language such as Java. This approach and its problems have been explained in Section I (and Figure 2). In Record & Replay, a tester runs the test script manually at first and the tool records the underlying implementation specifics of the AUT. The tool then replays these recorded steps. Record & Replay approach has similar drawbacks as the Programming approach as it uses the underlying code of the AUT for automation.

Image recognition based automation tools allows a tester to write a test step with the web element represented as an image. The tester first captures the image of the web element as displayed in the GUI and uses it in the test script. To execute such a test step, the tool locates that portion of the GUI which is similar to the image provided by the tester. This identification is done based on the intensity of the pixels (i.e. the code representing the pixels). Sikuli [5] is a popular tool using this paradigm. Figure 3 shows a test step in Sikuli.

The image recognition based method identifies web elements through a pixel to pixel comparison between two images. Such techniques have been found to take more time to create [8] since specific images to interact with every element needs to be taken (for example, selecting a value from a dropdown needs two images to be taken – an image to click on the dropdown and a second to choose the value. See Figure 5 depicting this case). Image recognition based techniques have also shown to be extremely brittle and needs significant maintenance [8] as simple changes to the AUT (such as a change in the web browser resolution or a change in the background color of a web page) can break the automation script.

In contrast, our method refers and identifies a web element based on how it is described in the textual test script rather than identifying based on the image of the web element (see Figure 3 for a comparison).

## III. TEST AUTOMATION USING COMPUTER VISION

In this section, we will provide the four use-cases depicting the advantages of our approach over the programming and image recognition based methods.

### A. Use Case 1: Author test scripts in a simple language without requiring implementation details of the AUT

Our approach allows a tester to author test scripts in an English-like language. Particularly, the test script needs no implementation details and needs no images of web elements. Figure 3 shows an example. Here, to click the 'Search button' on the webpage, the step needed in our method is the textual description of the intended action. In comparison, Selenium WebDriver needs the code to identify the button and Sikuli needs the image of the button.

At the time of execution, our method identifies the 'search button' from the GUI and performs the action.

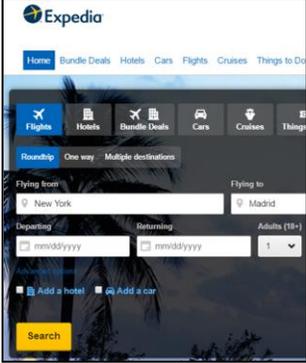
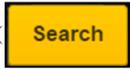

| | |
|---|---|
| | Scenario: to click the 'Search' Button |
| | Selenium WebDriver:<br>`WebElement searchButton = driver.findElement(By.id("srch-btn"));`<br>`searchButton.click();` |
| | Sikuli: Click ( [Search] ) |
| | Automation through Computer Vision<br>`When I click "Search" Button` |

**Figure 3. Automation test script to 'click a button'. In our method, the test step is written in descriptive text.**

## B. Use Case 2: Automate non-web based GUI applications

Many test scenarios span web & non-web GUIs. Consider the case of uploading a file where the GUI is rendered by Windows and not by the browser. There are no direct ways to interact with such GUIs through web-based automation systems such as Selenium WebDriver.

Figure 4 shows an example. In our method the test step is a description of the action intended and is the same for web & non-web GUIs. Conceptually, our method can work on systems with web-style GUIs on Desktop applications, native mobile applications, etc.

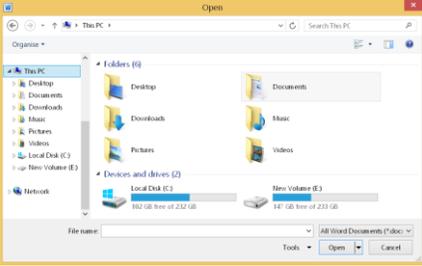

| | Scenario: upload a file |
| --- | --- |
| | Selenium WebDriver **Cannot be done.** |
| Sikuli: Click ( [Open ▼] ) | |
| Automation through Computer Vision **When I click "*Open*" Button** | |

**Figure 4. Automating non-web based GUIs.**

## C. Use Case 3 : Use visual relationships in Test Scripts

The visual relationship between different web elements can be exploited when test scripts are authored in our method. Consider the example of Figure 5. Note that there are two web elements – a label (with content as 'Title') and the dropdown. There need not be any relationship between these elements in the underlying code, however, visually there is a clear relationship.

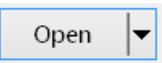

| | Scenario: select the Title of the person |
| --- | --- |

Selenium WebDriver

```
WebElement element =
driver.findElement(By.id("seltitle"));
Select dropDown = new Select(element);
Dropdown.selectByVisibleText("Mr");
```

Sikuli  Click ( [Title * Select ▼] )
        Click ( Mr. )

Automation through Computer Vision
**When I select "*Mr*" from dropdown "*Title*"**

**Figure 5. Refer web elements through visual relationships (e.g. the dropdown is referred by the label).**

A human would associate and refer to the dropdown as 'the Title dropdown'. However, this relationship cannot be used in the code written for Selenium. The Selenium code uses the identifier for the dropdown ('seltitle') and does not use the label component at all.

In our method, the dropdown that is associated with the text of 'Title' is deciphered visually. Such visual relationships are also useful in case of tabular data (where the perceived heading of a column can be referred to).

## D. Use Case 4: Resilient to changes

This case shows the effect of changes to the applications. In Figure 6, the relative position of the 'Title' label and the dropdown has been changed (we have highlighted the change in the example). Image based recognition techniques like Sikuli would fail in such cases. Further, if any underlying code has changed as well (for example changes to the identifier), the Selenium code has to be updated.

However, in either case, the automation script of our method need not be updated.

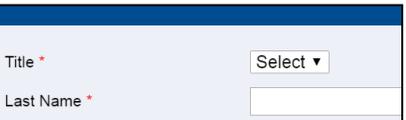

| | Scenario: select the Title of the person |
| --- | --- |

Selenium WebDriver

```
WebElement element =
driver.findElement(By.id("seltitlenewID"));
Select dropDown = new Select(element);
Dropdown.selectByVisibleText("Mr");
```

Sikuli  Click ( [Title * Select ▼] )  Click ( Mr. )

Automation through Computer Vision
**When I select "*Mr*" from dropdown "*Title*"**

**Figure 6. A change in the identifier of the web element requires updates to the Selenium code. A change in the visual representation requires a change in the Sikuli code. However, no changes are needed in our method.**

## IV. SOLUTION APPROACH

We have developed an initial solution approach tackling a limited set of web elements – buttons, textboxes, links and dropdowns. The solution approach is in Figure 7. However, we expect our approach to evolve as we work further.

Our solution approach is integrated into the Eclipse IDE. The Tester creates test scripts through a Domain Specific Language (DSL). We have built a DSL using English like statements on the xBase [2] framework. The DSL allows the tester to author test steps in a constrained natural language. The constrained natural language is easy to understand and yet allows a definite interpretation when parsed. We chose this approach over parsing an unconstrained natural language

(as attempted in [6]) since the DSL is not prone to ambiguities that are inherent in unconstrained natural language. A view of a few test steps written in the DSL is shown in Figure 8. Note that the keywords defined in the DSL constitute the actions (like '`I go to the URL`') and the test data or the element labels are in open ended text (shown in *blue*). Usage of xBase provides 'proposals' where a tester can choose the command from a list (and does need not to memorize the commands of the DSL).

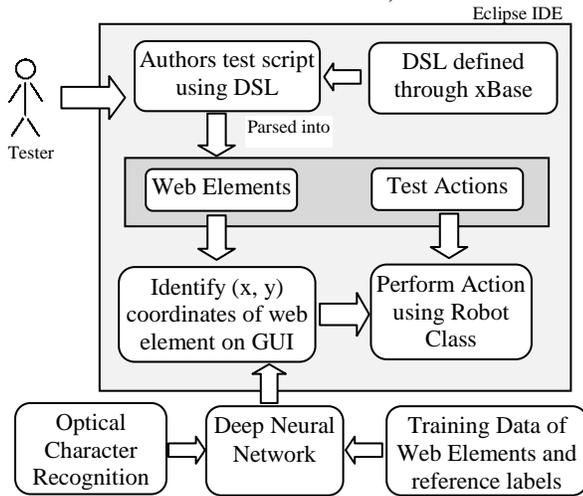

**Figure 7. Solution Approach of our method.**

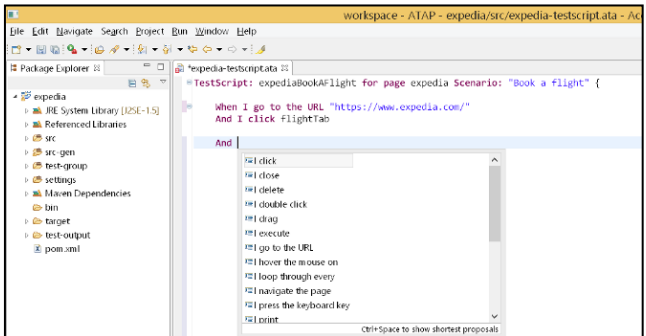

**Figure 8. Authoring a test script using our DSL.**

Once the test script has been authored, it is parsed to separate the web elements and the actions. For example, the test step of "`Enter 'New York' into the textbox 'Flying From'`" is broken into the web element of a textbox with label 'Flying From' and the action of "Enter 'New York'".

At this juncture, the web element is in textual notation. Our next task is to identify this web element in the GUI of the AUT. For this, a screenshot is taken and an Optical Character Recognition (OCR) system is used to transcribe all the text in the screen. We then identify the web element as textually described in the test step using a deep neural network. The neural network provides the visual (x,y) co-ordinates of the web element on the screenshot. The neural network has been trained on manually annotated data to identify web elements using labels. Further details on the neural net and the training are provided subsequently.

Once the co-ordinates of the web element are found, the required action is performed using the Java Robot Class. The Robot Class allows programmatic control of the mouse (to perform actions like click) and the keyboard (to perform actions like entering a text). The execution of the test script is complete at this juncture.

The computer vision approach to identify the web elements forms the fundamental novelty of our approach. We have used a deep neural network based on the work of [1]. This uses a Convolutional Neural Network (CNN) with a Long Short-Term Memory network (LSTM). We created a small amount of training data (about 2200 annotations) where the web elements were annotated manually with their reference labels. The neural network was trained on this data.

We have been able to perform limited testing and present some qualitative results. We found that the approach of using the DSL to author English-like test scripts works extremely well and the testers are able to adapt to it quickly. The parsing of the test scripts into web elements and actions is done well as well. The identification of the web elements through the neural network showed moderate results. However, we believe there is significant promise in the concept and we are actively extending our solution (with more training data).

## V. CONCLUSION

Automation in software testing has seen limited adoption in practice. The problems in the current automation methods are manifested in the disconnect between the way the automation is done versus the way manual testing is done.

In this paper, we propose to bridge this gap by building a machine that can test software like a human. The method allows a tester to author a test script in an English-like language without referring to any implementation specifics of the AUT (making it easy to create & maintain). Our method identifies the web element referred in the test script through recent advances in computer vision. We have built a Convolutional Neural Network using a Long Short-Term Memory network which has been trained on a manually annotated data of web elements with labels.

Initial results are promising and we believe our method has the potential to significantly advance test automation.